\documentstyle[11pt]{article}
    
\begin{document}

\begin{flushright}
BRX TH--457
\end{flushright}

\begin{center}
{\Large\bf Nonrenormalizability of (Last Hope)
D=11 Supergravity, \\with a Terse Survey of
Divergences in Quantum Gravities}

\vspace{.4cm}

{\sc S. Deser}\footnote{\tt deser@brandeis.edu}\\

\vspace{.3cm}

{\em Department of Physics} \\
{\em Brandeis University, Waltham, MA 02454, USA} 

\vspace{.2cm}
\end{center}

\begin{quote}
{\bf Abstract}:
Before turning to the new result that D=11 supergravity
is 2-loop nonrenormalizable, we give
a very brief history of the ultraviolet problems
of ordinary quantum gravity and of 
supergravities in general D.
\end{quote}

\addtocounter{section}{1}

\noindent{\bf 1. Introduction}

The organizers have asked me to preface my presentation
of the 2-loop nonrenormalizability of D=11 supergravity \cite{001}
(SUGRA) with a historical survey of the ultraviolet problems in Einstein
theory and in lower dimensional SUGRAs.  I am happy to comply, as this
will help in understanding why I called D=11 SUGRA the last hope:
it was the only local Quantum Field Theory that contains general
relativity whose non-renormalizability properties had not yet been
established.  Then, as an introduction to our results, I will
motivate the analysis and its implications.  Naturally,
I will be brief both in my survey and references, given the space
and time constraints of this conference; by a happy coincidence,
the new work with D. Seminara appears in the synchronous
issue of Phys.\ Rev.\ Lett. \cite{002}.

\renewcommand{\theequation}{2.\arabic{equation}}
\setcounter{equation}{1}

\noindent{\bf 2. The Ultraviolet Problems of General Relativity}

\noindent This is a subject with little prehistory, aside from an
old remark of Heisenberg that theories with (positive)
dimensional coupling
constants would be ill-behaved at high energies, one that
 will be amply
borne out by the following considerations.

We will be working throughout in the standard perturbative 
formulation of GR about a flat space vacuum, expanding the metric
in powers of $\kappa h_{\mu\nu} \equiv g_{\mu\nu}-\eta_{\mu\nu}$.
This displays the D-dimensional Einstein action,
$$
I_E = \kappa^{-2} \int  d^Dx \: \sqrt{-g} \: R \eqno{(2.1{\rm a})}
$$
as an infinitely self-coupled QFT,
$$
I_E = \int d^Dx \: (\partial h)^2 \sum^\infty_{n=0}
(\kappa h)^n \; ,
 \eqno{(2.1{\rm b})}
$$
homogeneous in second derivatives.  The quadratic $(n=0)$ terms
describe the usual $q^{-2}$ propagator, the cubic $(n=1)$ the 
3-point vertex describing the lowest self-interaction of the 
$h$-field with its own stress-(pseudo) tensor 
$T_{\mu\nu}(h) \sim (\partial h)^2$, and so on.
The dimensions of $\kappa^2$ are, by (2.1a), $\sim L^{D-2}$,
which will be important in what follows.

Let us dispose of some 
special cases.  The D=3 theory is well-known to have no 
propagating modes \cite{003} and correspondingly it is
finite \cite{004,005} despite having a dimensional 
$\kappa$.  For completeness, we mention that at D=3 there is
a third derivative order but unitary 
model with local excitations, topologically massive gravity 
\cite{006}, for which the verdict is still not
known \cite{007}.  For D=2, the Lagrangian is the Euler
density and no purely gravitational Einstein theory exists.
We will also not deal with higher derivative
order theories, with terms
quadratic in curvatures; they are renormalizable 
but at the price of ghost modes, since they typically have
propagator denominators $\sim p^{-4}$ or $(m^2p^2 + p^4 )^{-1}$.
Viewed as fundamental actions, they are thus really regulator
terms; if, instead, the $R^2$ parts are themselves
viewed as effective perturbative
additions to the Einstein action, they cannot be used to
``improve" the Einstein propagator.

Consider the simplest one-loop self-energy diagrams 
using dimensional regularization -- which organizes divergences
most transparently.  Since the vertex and propagator are of
reciprocal powers $(p^2,p^{-2})$ in momentum, and since
each external line $\sim (\kappa h)$ must acquire two 
$p$'s to maintain gauge invariance by becoming a curvature,
we see that each loop order acquires potential
divergent contributions proportional to
\begin{equation}
\sum^\infty_{\ell = 1} \left[
 \Delta I_\ell = \int d^Dx \:
R^{D/2}  (\kappa^2 R^{D/2-1})^{\ell -1} \right]
\end{equation}
where $\ell$ represents the loop order and $R$ stands for
generic curvatures or two covariant derivatives.
[In odd dimensions there can only be even loop divergences.]
Thus, in principle, unless there is a reason for the 
vanishing of an infinite number of coefficients,
the theory loses predictability at all orders -- it is 
non-renormalizable.    The above result
is for pure gravity.  When the latter is coupled to normal
matter, there will be further counterterms: graviton loops
will generate matter-dependent ones, while 
matter loops also contribute curvature-dependent
divergences.  [Recall that (boson, fermion) propagators go
as $(q^{-2},q^{-1})$ and their minimal coupling to gravity
through their stress tensors go as $(q^{+2},q^{+1})$, 
expressing the universality of gravitational couplings.] 
This will lead to an expansion analogous to (2.2), involving
powers both of $R$ and of $\kappa^2 T_{\mu\nu}$--like terms.

What are the concrete outcomes of this general framework?
In the early seventies, a transparent algorithm
to calculate generic one-loop graphs was presented
\cite{009} and exploited to make several important 
conclusions for D=4:  In pure gravity, 
while the quadratic curvature's coefficients in (2.2) 
do not vanish, this is actually
easily remedied by a (divergent) field redefinition because 
the Gauss--Bonnet identity, $\int d^4x 
[R^2_{\mu\nu\alpha\beta} - 4 \: R^2_{\mu\nu}
+R^2] = 0$, made these terms proportional to the field equation,
$\delta I_E/\delta g_{\mu\nu} \sim 
\int d^4x \, G_{\mu\nu}X^{\mu\nu}$.  
The gravity-scalar field system does contain infinite and 
non-removable terms, {\it i.e.}, they are not 
proportional to the field equations, $\Delta I \neq
\int  d^4x \, (G_{\mu\nu} -\kappa^2 
T_{\mu\nu}(\varphi )) X^{\mu\nu}$.  
In the wake of these results, 
other relevant matter couplings, including fermions,
Yang--Mills and QED were systematically explored \cite{010}.
In all cases, their gauge or 
fermionic character was of no help; they all 
failed the one-loop test.  Beyond one loop, pure GR
can exhibit invariant counterterms proportional to
cubic and higher powers of the Weyl tensor that do not
vanish on-shell.  
The much harder job of explicitly calculating that 
pure GR failed at 2-loop order was successfully undertaken 
in \cite{011,012}; the coefficients of the 
$R^3_{\mu\nu\alpha\beta}$ counterterms were indeed non-zero.

Because the above failures of GR were at the perturbative level, 
I should note for balance
that already in the late '50s it was suggested
that  GR might be a universal regulator
for all QFT in some nonperturbative way, 
a development that was to materialize much later and 
in an unexpected way through strings.  
Indeed, that closed strings are both finite and 
contain $s$=2, $m$=0 excitations, providing an acceptable
quantum version of GR, is an essential part of their 
(and their successors') central role.

\renewcommand{\theequation}{3.\arabic{equation}}
\setcounter{equation}{1}

\noindent{\bf 3. D$<$11 Supergravities}

One of the hopes following the discovery of D=4 supergravity in 
1976 was that, despite also describing gravity plus ``matter",
it would be more convergent than the systems described earlier.
The reason was of course the additional supersymmetry
(SUSY), which combined spin 3/2 ``matter" and gravitons into 
a single (super) multiplet.  Indeed, there was both one- and
two-loop improvement: The 
one-loop infinities were precisely as in pure gravity
(rather than as in gravity coupled to ``ordinary" matter),
{\it i.e.}, field-redefinable arrays proportional to the
field equations of supergravity, {\it e.g.},
$\Delta I_1 = \int d^4x (G_{\mu\nu} - \kappa^2
T_{\mu\nu} (\psi )) X^{\mu\nu},$
where $\psi_\mu$ is the vector-spinor companion of the graviton,
and the two-loop term was absent altogether due to supersymmetry,
there being no companion to $\kappa^2R^3_{\mu\nu\alpha\beta}$.
Far more important however, \cite{014}, three-loop invariants 
did exist for the N=1 and higher models. 
Since these will have their counterparts in D=11, let me sketch
their form.  It is known in the simpler global SUSY case that
the system's stress tensor $T_{\mu\nu}$, supercurrent
$J_\mu$ and chiral current $C_\mu$ form a 
supermultiplet from which the SUSY invariant
$$
I \sim \int d^4x [T^2_{\mu\nu} -i \bar{J}_\mu \not\partial J^\mu
+ \textstyle{\frac{3}{2}}C_\mu \Box C^\mu ]  \eqno{(3.1{\rm a})}
$$
is constructible.  Now as is well-understood, there is no invariant
stress-tensor for the gravitational field itself, and the
supercurrent is of higher derivative (and 3-index) order,
$J \sim (Rf)$ where $f_{\mu\nu} =
D_\mu \psi_\nu - D_\nu \psi_\mu$ is the field strength of the
potential $\psi_\mu$.  It is thus natural to seek a 
higher derivative analog of $T_{\mu\nu}$, and there is one,
namely the well-known Bel--Robinson tensor 
$B_{\mu\nu\alpha\beta}$ defined uniquely in D=4 according to
\setcounter{equation}{1}
\begin{equation}
B_{\mu\nu\alpha\beta}=R^{\rho\ \sigma\ }_{\ \mu\ \alpha} 
R_{\rho\nu\sigma\beta}+R^{\rho\ \sigma\ }_{\ \mu\ \beta} 
R_{\rho\nu\sigma\alpha}-\textstyle{\frac{1}{2}} g_{\mu\nu} 
R_{\alpha}^{\ \rho\sigma\tau}R_{\beta\rho\sigma\tau}. 
\end{equation}
On-shell $(R_{\mu\nu} = 0)$ $B$ is totally symmetric,
(covariantly) conserved and traceless and there is again an 
invariant like (3.1a) namely
$$
\Delta I_3 = \kappa^4 \int d^4x 
[B^2 -i \bar{J} \not\partial J
+ \textstyle{\frac{3}{2}} C \Box C ]  \eqno{(3.1{\rm b})}
$$
where I have omitted all indices ($C$ is a chiral current
bilinear in $f_{\mu\nu}$) and added the $\kappa^4$ to show
eligibility of $\Delta I_3$ as a 3-loop counterterm.
The existence of similar
invariants for N$>$1 was soon confirmed as well
\cite{015} and their full nonlinear completions 
were found by superspace methods that are, unfortunately,
not available in our D=11 world. Surprisingly, 
it was even possible to learn something about the
actual coefficients of such counterterms:  For the non-maximal
1$\leq$N$<$8 models, where covariant superspace quantization
is possible, it was concluded \cite{new14} that they
are uniformly non-vanishing.
For the special maximal N=8 case, very beautiful recent
work \cite{017} (to which we shall return for its impact
on D=11) suggests that, while the three-loop coefficient may
vanish, its 5-loop(!) coefficient probably does not.

To summarize, then, the D=4 SUGRA situation is essentially that
the one-loop counterterms are accidentally protected by the
supersymmetric extension of the Gauss--Bonnet identity,
{\it i.e.}, they vanish on shell and hence are safe,
whatever their coefficients.  At two loops, whose bosonic
part must start as $\Delta I_2 \sim \kappa^2 \int
d^4x [R^3_{\mu\nu\alpha\beta}]$, there exists no SUSY
completion; this is the improvement over
pure gravity, where this $R^3$ term is both allowed and has
nonzero coefficient.  On the other hand, at 3 loops there
is indeed the allowed N=1 term we have displayed in (3.1b)
and no miracles protect its coefficient or that of its
N$<$8 counterparts; even for the maximal N=8, 
five loops prove fatal.

I will not repeat this general SUGRA argument for rising
dimensions, 4$<$D$<$11, (again even and odd ones differ
a bit) except to say that it is possible to construct
SUSY counterterms explicitly where a suitable superspace
formulation exists and that the construction and negative
conclusions of \cite{017} seem to apply.  
Instead, I now move to my main topic, D=11 SUGRA.

\renewcommand{\theequation}{4.\arabic{equation}}
\setcounter{equation}{0}

\noindent{\bf 4. D=11 SUGRA}

This section represents the work in \cite{002} to
which I refer for details.  In the period of rapid 
construction of D$>$4 SUGRAs following the
initial D=4 theory, it became clear
from the mathematics of graded algebras \cite{018}
that if one wanted these models to obey the following
four criteria:
inclusion of s=2, $m$=0 excitations, {\it i.e.}, the
graviton (of which Einstein theory is the essentially 
unique local model); only one such graviton;
no massless excitations with spin exceeding 2, and
only one timelike dimension,
then there is a highest dimension, D=11, for SUGRA.  This
theorem can be understood in various ways, in particular
from the requirements that the numbers of boson/fermion
excitations must match.  The $s<5/2$ requirement
is connected with the fact that massless fields with 
$s\geq 2$ do not consistently interact with gravity itself
\cite{019}.  The D=11 theory \cite{001} is indeed
a very special one, even in the universe of SUGRAs. 
Some examples: 
it allows only 
N=1, so ``maximal = minimal"; there is no
SUSY ``matter" in D=10 to provide a source; while 
(negative) cosmological
constants can be included in lower-dimensional 
SUGRA, (unbroken) D=11 SUGRA is also unique in forbidding 
a cosmological term \cite{020}.

Before going into 
the description of our construction and its implications
for the theory's nonrenormalizability, let me turn to our
motivations. They are actually twofold.  The immediate aim
was to construct on-shell non-vanishing local invariants,
as potential counterterms.   Now this was already a very
nontrivial task, because in this theory there is no 
technology to test the SUSY of, let alone construct,
candidate invariants.  Even assuming these would be likely to
exist, this was the last remaining SUGRA model and  (with
its mysterious other uniqueness properties) the only one 
with a chance at staving off infinities.
Besides, having been reenthroned (from its previous anomalous 
position in the old D=10 string world) as  
a basic cornerstone/QFT limit of M-theory, it is not only 
``there", but of prime interest!  Secondly, quite
apart from renormalizability, any higher invariant that could
be constructed would automatically be a candidate 
(finite) correction term in some M-theory expansion and
as such be a window on that mysterious region, precisely in
the same way that D=10 string models yielded (also finite)
corrections to the D=10 supergravities.  So much for why --
now we need the how.

To know where to focus, let us first re-count dimensions,
which is best done by looking at the SUGRA action.  I will
only write the purely bosonic part here, because that is 
all we will (fortunately!) need:
\setcounter{equation}{0}
\begin{equation}
I^B_{11}= \! \int \! d^{11}x \left [-\frac{\sqrt{g}}{4\kappa^2} R(g)
-\frac{\sqrt{g}}{48} F^2
+\frac{2\kappa}{144^2}\epsilon^{1\cdots 11}
F_{1\cdots}F_{5\cdots}A_{..11}\right],
\end{equation}
This is not the place for a review;
let me just recall that in addition to the graviton,
there is a 3-form potential $A_{\mu\nu\alpha}$
with associated field strength
$F_{\mu\nu\alpha\beta} \equiv \partial_{[\beta}
A_{\mu\nu\alpha ]}$ and a cubic Chern--Simons (CS)
term in which the 11 indices of the epsilon symbol
are saturated by two $F$'s and one $A$.  The dimensions 
of $\kappa^2$ are here
$L^{+9}$; note the explicit $\kappa$ also in the CS
part since  
$A\sim L^{-9/2}$.  Among the fermionic terms, there are 
non-minimal ones $\sim \bar{\psi} R\psi, \;
\bar{\psi}F\psi$ but we can avoid this whole
sector here.  It is clear, because dimension is odd, that
no 1-loop $\sim \kappa^0 \int d^{11}x$ candidate
$\Delta I_1$ exist -- one cannot make gravitational
scalars with odd numbers of derivatives, except irrelevant
(because parity-violating) $\epsilon \, R^n \partial R^m$
terms.  At 2-loop order, 
\begin{equation}
\Delta I_2 \sim \kappa^2 \int d^{11}x \, \Delta L_{20}
\end{equation}
where $\Delta L_{20}$ has dimension 20.

To construct the leading purely gravitational term here
requires, at first sight, $R^{+10}$.  While such terms
are undoubtedly possible and present, they are
impracticable to obtain.  Related to this is the question of
regularization scheme.  If we use some dimensional 
(energy) cutoff, it alters the allowed dimension of
candidate $\Delta L$'s.  Instead, the dimensional
regularization we use here uniformly has logarithmic
cutoff so all $\Delta L$ are $\Delta L_{20}$'s.
But all we have to do is to present one that exists
and show that its coefficient fails to vanish; if we
can accomplish that as we (essentially, see below)
will, other possible regularization schemes are 
irrelevant.  Coming back to finding a more tractable 
$\Delta L_{20}$, we recall that a curvature is
dimensionally equivalent to two covariant
derivatives $D_\mu$, which means that candidate terms are
schematically of the form $\Delta L_{20} \sim
\sum^{10}_{n=4} R^n \, D^{2(10-n)}$.  We start at
$n$=4 because clearly the lower $n$'s are either
(like $R^3$) not parts of super-invariants or are leading
order trivial (like $R^2$ which obeys Gauss--Bonnet
to quadratic order in $h_{\mu\nu}$).  Thus, our
lowest possible choice is $\Delta I_2 \sim \kappa^2
\int d^{11}x [R^4 D^{12} + \ldots ]$ where the
ellipsis represents the SUSY completion, if any.
How to find a suitable candidate in absence of any
guiding super-calculus?  Our procedure was the
following.  As was also recognized in \cite{017},
there is certainly one on-shell nonvanishing
lowest order SUSY invariant that starts out quartic
in $h_{\mu\nu}$, and that is the tree-level 4-point
scattering amplitude generated by the D=11 action
(4.1) itself.  It has the enormous advantage that,
since there are no loops, and SUSY transformations are
linear at our level, the purely bosonic terms are
guaranteed to be part of the overall SUSY invariant that
is the total 4-point amplitude.  However, it presents
two {\it a priori} obstacles:  First, we want a local
invariant, whereas the amplitude has a denominator, from
virtual particle exchanges; can we extract a local 
but still SUSY residue?  We can, simply
because each term in the amplitude is in fact 
proportional to the product $(1/stu)$ of the Mandelstam
variables.  Second, we want to have 12 explicit
derivatives in the $R^4$ and other terms; can those 
be inserted without losing SUSY or having everything
vanish on-shell?  The answer is again yes, {\it e.g.}, by
further multiplication with $(stu)^2$ or 
$(s^6 + t^6 + u^6)$, after the initial $stu$ one.

Now we have cleared the decks for the actual computation.
It consists of applying the Feynman rules in terms of the
propagators and
vertices needed up to 4-point level.  Let's review the
ingredients.  Expanding the action (4.1) about flat space
gives us first the free particle propagator terms,
symbolically,
\begin{equation}
I^{(2)}_{11} \sim \int d^{11}x (h D_g^{-1} h + A
D_F^{-1} A) \; ,
\end{equation}
where ($D_F$, $D_g$ are the respective free 
($\sim \Box^{-1})$ propagators.
The cubic terms are of three types.  First, the purely 
topological (metric-free) CS term furnishes the 3-form 
self-coupling vertex $\kappa \epsilon FF A \equiv C_F
 \cdot A$ where the three-index current $C_F \sim 
\kappa (\epsilon
FF)$ is both gauge invariant and identically conserved. 
The other relevant vertex involving 
$F$ is its gravitational coupling
$\kappa h_{\mu\nu} T^{\mu\nu}_F$ where
$T^{\mu\nu}_F \sim FF$ is the form's stress-tensor.  Finally
the purely gravitational contribution is of the same
$\kappa h_{\mu\nu} T^{\mu\nu}(h)$ form, where 
$T^{\mu\nu}(h) \sim (\partial h \partial h)^{\mu\nu}$ is
the (gauge-variant) 
stress pseudo-tensor of the gravitational field
itself.  This term, upon contracting one of its graviton
lines with that of another, yields the graviton-graviton 
exchange $\sim \kappa^2 T^{\mu\nu} (h) 
D^g_{\mu\nu\alpha\beta} T^{\alpha\beta} (h)$ 
just as self-contracting
the $A\:C_F$ term contributes $C_F \, D^F \, C_F \sim
\kappa^2 F^4$ to form-form scattering, 
as does the $h T_F - h T_F$
graviton contraction.  However, the four-graviton
term is not even abelian gauge
invariant by itself but requires inclusion of 
the local quartic 4-point vertex 
$\kappa^2 hh\partial h \partial h$ to restore it.
Besides the above pure $h^4$ and $F^4$ amplitudes, there
are mixed terms:  If  we contract
the form field in $h T_F - hT_F$, we get $\sim F^2R^2$,
form-graviton Compton scattering.  Finally we can get a
form-graviton bremsstrahlung term  $\sim RF^3$ when
contracting the $AC_F$ and $h\, T_F$ vertices (only form
exchange contributes here).  Single-form creation 
$\sim R^3F$, is of course forbidden.

It is a straightforward, if index-intensive, procedure to
perform these calculations.  The worst, graviton-graviton
scattering has fortunately been done earlier in arbitrary
dimension \cite{021} which is a very useful check;
the answer is as in D=10, namely it is proportional to
the famous lowest string correction to the Einstein action
in D=10:
\begin{equation}
L_g = stu \: M_g \sim t_8 t_8 \: RRRR
\end{equation}
where $t_8$ is a constant 8-index tensor; it has a piece
proportional to the D=8 Levi-Civita symbol $\epsilon_8$  
and hence the 
$\epsilon_8\epsilon_8$ part of $L_g$, proportional to
the D=8 Gauss--Bonnet invariant, cannot be seen at our 
lowest, ${\cal O} (h^4)$, order (at its lowest order, the 
G--B term of any dimension is a total divergence in all
D).  Remember that we are always on (linearized) shell,
so that the letter $R$ really means the Weyl tensor;
also, we have emphasized that multiplication of $M$ by
$stu$ yields the local scalar $L$.

Let me now discuss more explicitly our bosonic 
component of the full SUSY invariant; as mentioned
earlier, it is interesting in its own right as the 
correction to the D=11 SUGRA action from M-theory, of
which we mostly know that it contains SUGRA
as the local limit.  Below, I will summarize the appearance
of the various ``localized" on-shell 4-point
amplitudes.  They are to be multiplied by the required
twelve derivatives, say with $(stu)^2$ to make dimension
20.  Schematically (see [2] for details), with $B$ a
Bel--Robinson-like curvature quadratic and $F$ always
appearing with a gradient,
\begin{equation}
\Delta I^B_2 (g,F) = \kappa^2 \int d^{11}x 
[B^2 + (\partial F)^4 + B(\partial F)^2
+R (\partial F)^3] \; .
\end{equation}
Although everything looks very coordinate invariant, these terms
are only accurate to lowest, 4$^{\rm th}$, order in the
combined $R$'s and $F$'s, and of course are to be 
supplemented by fermion-dependent terms that we
do not write down.  Nevertheless linearized SUSY is
guaranteed by our construction.

What about the coefficient, is it non-zero?  As I
mentioned, a powerful tool for answering this question
was provided by the amazing 
correspondence between SUGRA and Super-Yang--Mills 
models established and exploited in \cite{017}
to obtain otherwise ``impossible"
results.  The only catch is that SYM is only defined for
D$\leq$10; however one can argue, quite convincingly,
that the results as provided really do not depend directly
on D, and extend analytically
also to D=11.  At any rate, once their D=11
extension has been made, it is of course independent of
its origin and it may not be all that difficult to verify
intrinsically; thus, the strong odds are against even this
maximal theory, already at leading
possible level. 

\setcounter{equation}{0}

\noindent{\bf 5. Summary}

We have tried to outline in a very short space, the history
of the ultraviolet problems of any QFT that contains 
general relativity, but (to preserve unitarity) no higher
derivative kinetic terms.  
These models include pure gravity in
any D$>$3, where the problems arise uniformly at 2-loop
level because the 1-loop terms, while still infinite are
accidentally harmless \cite{009}.  For D=4 in particular
the coefficient of the 2-loop counterterm has been
explicitly and reliably shown not to vanish \cite{011,012}.
For gravity plus lower spin $(0 \leq s \leq 1)$ matter
the one-loop terms are already known to be non-zero,
also explicitly \cite{010}.  All these problems are traceable
to the positive dimensionality of the gravitational
coupling constant, $\kappa^2 \sim L^{D-2}$.

The addition ``gravity-matter" symmetry that is the 
hallmark of supergravity was shown early on, for the
D=4, N=1 model, to improve things but only
marginally: there the danger comes from 3 loops onward
\cite{014}; things are not improved by higher N
\cite{new14} although it may be \cite{017} that the
maximal, N=8, model is only destroyed at
5-loop order.  Increasing dimension is no help either,
and indeed our main new result is that the most
unique and extreme, D=11, SUGRA
is also sick already at minimal 2-loop order, despite
its relation to the M-theory unification of string
theories.  More specifically, a
local SUSY invariant counterterm was
constructed from the 4-point tree-level
scattering amplitude generated by the D=11 action. 
Together with the compelling argument of \cite{017}
that its coefficient is non-null, this conclusion is
all but inevitable. Hence Heisenberg's curse on every
QFT that includes gravity holds universally and forces
us beyond locality, to strings and their generalizations.

\vspace{.75cm}


\noindent {\bf Acknowledgements}: This work was supported by
the National Science Foundation under grant PHY93-15811.
I am grateful to my D=11 coauthor D.\ Seminara, as well as to
Z.\ Bern and L.\ Dixon, for useful comments.

\vspace{5mm}
\newpage


\end{document}